\documentclass[sigconf]{acmart}

\usepackage{url}
\usepackage{breakurl}
\usepackage{balance}

\copyrightyear{2018}
\acmYear{2018} 
\setcopyright{iw3c2w3}
\acmConference[WWW '18 Companion]{The 2018 Web Conference Companion}{April 23--27, 2018}{Lyon, France}
\acmBooktitle{WWW '18 Companion: The 2018 Web Conference Companion, April 23--27, 2018, Lyon, France}
\acmPrice{}
\acmDOI{10.1145/3184558.3191613}
\acmISBN{978-1-4503-5640-4/18/04}

\begin{document}
\title{MMM: May I Mine Your Mind?}

\author{Diego Sempreboni}
\affiliation{%
  \institution{King's College London}
  \streetaddress{P.O. Box 1212}
  \city{London}
  \state{United Kindgom}
  \postcode{43017-6221}
}
\email{diego.sempreboni@kcl.ac.uk}

\author{Luca Vigan\`{o}}
\affiliation{%
  \institution{King's College London}
  \streetaddress{P.O. Box 1212}
  \city{London}
  \state{United Kindgom}
  \postcode{43017-6221}
}
\email{luca.vigano@kcl.ac.uk}

\renewcommand{\shortauthors}{D.~Sempreboni and L.~Vigan\`o}

\begin{abstract}
Consider the following set-up for the plot of a possible future episode of the TV series Black Mirror: human brains can be connected directly to the net and MiningMind Inc. has developed a technology that merges a reward system with a cryptojacking engine that uses the human brain to mine cryptocurrency (or to carry out some other mining activity). Part of our brain will be committed to cryptographic calculations (mining), leaving the remaining part untouched for everyday operations, i.e., for our brain's normal daily activity. In this short paper, we briefly 
argue why this set-up might not be so far fetched after all, 
and explore the impact that such a technology could have on our lives and our society.
\end{abstract}





\maketitle

\section{What would you be prepared to do for a free Apple Watch?}

Everything has a price. Even us. Even you. And when it appears that something or someone doesn't have a price, well, then it could be that you are the price. For instance, what would you be willing to do to get the latest model of an Apple Watch (or of an iPhone or of a Samsung Galaxy, etc.) for free? What is your price? 

The lure of easy money is intrinsic in human nature. Since the dawn of time, human beings have kept inventing novel methods to get rich, legally or illegally, sometimes coming to a compromise even with their own principles, sometimes compromising themselves.

The widespread of the Internet and the progress of web technologies have only increased the number of ways to make money easily. One of the most recent examples is provided by the \emph{fake news and clickbait} phenomenon. Fake news are false stories that at first glance might appear to be real news, spread on the Internet or using other media, usually used to stoke the clickbait, which is an eye-catching link on a website that encourages people to read on.

The purpose of fake news is propaganda or profit, and sometimes both at the same time. As a timely example of the former, consider the on-going investigation on the influence and interference of the Russian government in the 2016 United States elections. For the latter, observe that once an internaut has clicked on a catchy title and opened the fake news page, the owner of the website will earn a small amount of money based on the advertisement on that page. This means that, as the number of views increases, the  gain of the website's owner will also increase.

In the last couple of years, a large amount of websites have been created that have the sole purpose of deriving profit from the views thanks to the advertisement system even if this means exploiting and, in some cases, ridiculing and mocking human emotions~\cite{engadgetfake}. For instance, consider what happened in the aftermath of the terrorist attack in London in 2017 during which 6 people were killed and 50 were injured~\cite{bbc}. One of the pictures that became viral, and fueled the anger and hatred of the net, is that of a woman wearing a hijab quite nonchalantly walking past victims on Westminster Bridge while talking on phone. As later clarified by the photographer who took the picture, the woman had actually tried to help and in fact she released a statement soon after the attack to explain that: ``What the image does not show is that I had talked to other witnesses to try and find out what was happening, to see if I could be of any help, even though enough people were at the scene tending to the victims. I then decided to call my family to say that I was fine and was making my way home from work, assisting a lady along the way by helping her get to Waterloo station.''
Another fake news story that was shared millions of times is the report that the hurricane Irma that devastated the Americas reached the category-six storm... ignoring the fact that category-six hurricanes actually don't exist.

In order to maximize customer visits, the news relate to hot and trending topics, and use titles that appeal to feelings such as anger, frustration or morbid curiosity. Moreover, the authors of fake news take advantage of social media to spread the news and capture new viewers.
Actually, also social media make easy money just because we provide our information, along with our permission to use it, when we accept their conditions (often without reading these conditions at all). In a sense we, or to be more specific our information, have become the currency for Facebook, Instagram, and the like.
This is, of course, because human nature can be exploited easily: we tend to publish personal information without thinking about the consequences, we click on news that we believe to be true and we share them, fueling the online advertising market. Unfortunately, technology isn't yet able to help in this case: ad blockers and anti-tracking have been successfully applied to mitigate the nuisance of online advertisement, thereby lessening the gain of the owners of fake-news websites and thus disincentivizing the creation of such sites, but work is still ongoing on developing tools that can efficiently detect and block fake-news~\cite{conroy2015automatic, fakenewsmozilla, tacchini2017some, wang2017liar}.

However, a new approach has been recently adopted by website owners to increase their earnings.~\cite{coin}. This consists of a technology based on JavaScript but, taking advantage of hot and trending ideas such as blockchain and cryptocurrencies, could be seen as a direct consequence for allaying the successes of ad-blockers in blocking classical ads~\cite{wired,safari}. More specifically, some websites have adopted a new method called \textit{cryptojacking}~\cite{cryptodefinition, STOKELWALKER201816}, which is the hidden use of a computing device to mine cryptocurrency. This recent phenomenon started with some Torrent search engine websites~\cite{piratebay}, but subsequently several cryptojacking engines were found on different streaming websites~\cite{streamingcrypto} and even in some adverts on the popular YouTube platform~\cite{youtube}. 
This seems to be just the tip of the iceberg: cryptojacking has recently been used as a vector for wide-spread cyber-attacks (e.g., \cite{hackcampain,thousand}) and thousands of websites have been hacked with the goal of injecting cryptojacking engines.

Although (sometimes) cryptojacking may appear to be a nuisance and a burden for the internauts, its use has been willingly accepted in some cases, especially in situations in which the advantages the users can get overweigh the imposition. All of this, to make ``life'' easier, better and, mainly, free.

At the beginning of this section, we asked what you would be willing to do to get an Apple Watch for free. A question like this has been used by Vitality, a company that sells health and life insurances, in order to reward its customers~\cite{vitality,applewatch}. Vitality has launched the \emph{Vitality Active Rewards} program that aims to promote physical activity and better health for their customers, who can earn Vitality points by reaching some activity goals thanks to the tracking of their activities using a wearable device (e.g., an Apple Watch). The points are used on a monthly bases to discount the price of the wearable provided by Vitality. 
Hence, particularly engaged customers could actually earn their Apple Watch series 3 for free.

\section{A reward system}
This kind of marketing idea is the base of what is typically called a \emph{reward system}, where a customer has the chance to obtain a benefit providing in return a ``dual'' profit to those who provided the benefit. In the Vitality example, customers who keep fit will likely cost less to the insurance company (e.g., in terms of health-care costs) so that the company can reward those customers with a monthly discount for the watch. This is, in a sense, a technological equivalent of the discounts offered by some health-insurance companies to customers who verifiably 
quit smoking. 

Different kinds of reward systems exist, ranging, for instance, from more old-fashioned kinds such as companies distributing merit awards to the employees who achieve high productivity standards~\cite{kopelman1977merit}, to mobile applications that ask users to scan the bar code of some products in a shopping mall in order to be rewarded
with points that they can spend to claim prizes, e.g., \cite{checkpoints,shopkick}. Also the aforementioned streaming service can be seen as a reward system where the user has the chance to access a streaming platform but paying a price through advertisements or CPU resources.

Reward can take also other less tangible forms, such as those that have been dramatized in the movie ``Nerve'' and in several Black Mirror episodes (most notably, ``Nosedive'' S03E01).

\emph{Nerve}~\cite{nerveimdb}, a 2016 movie directed by Henry Joost from a screenplay written by Jessica Sharzer based on the novel by Jeanne Ryan, intertwines profit with \emph{popularity}. This is just the logical consequence of what is already happening: today, to be extremely popular in a social network means to have the chance to make money and new ``professional figures'' have emerged that are using popularity for their business (e.g., youtuber, influencer).
Nerve depicts a society where a famous virtual reality game is based on the idea of ``Truth of Dare''.
Add a pinch of social media with popularity and economic gain and we have the perfect reward system of our days. The Nerve application is a further step forward for this concept because the system dares the players to accomplish some challenges, rewarding them with amounts of money based on the difficulty of the challenge. So, the human is put on the front line much more than in other scenarios.

Black Mirror's episode \emph{Nosedive} (S03E01) depicts how social appearances become valuable: the number of likes received increases not only the sense of fulfillment and accomplishment of the individuals but also their recognition by the whole society, and thus their chance of climbing the social ladder.

In both Nerve and Nosedive, as well as in other Black Mirror episodes such as \emph{Hated in the Nation} (S03E06) and the special \emph{White Christmas}, humans are always online. Smartphones are the first mobile devices that allowed us to connect to the Internet in a wide-raging and uninterrupted way, but the trend is to invent new technologies that allow us to be connected even without a smartphone~\cite{without}. The idea underlying the \emph{Internet of Things} is based on this and we actually are one of those ``things''. However, it is not hard to imagine a future in which we will no longer need to bring with us a smartphone or a wearable like a smartwatch or a smartband. 

Several studies have been published on brain-computer interaction~\cite{Tan2010} and there are start-ups that are developing interfaces to connect the human brain directly to a computer~\cite{neuralink}, but it is not science fiction to imagine that soon we won't even need such interfaces as we will be directly connected to the Internet. In fact, some preliminary work has already been carried out~\cite{linesbrain}.

Our brain will be connected. Always connected.
Our brain will be always able to retrieve data from the network and process it. This means that our brain will have a direct channel to the Internet, so why not imagine traffic going in both ways? Why not, even without a full understanding of our brain, try to find a way to use part of it as a CPU, the same CPU we have in our laptops? 
This may be the business idea of a new company that we have called \emph{MiningMind Inc}.

\section{Miningmind Inc.}

Consider the following plot set-up for a possible future episode of Black Mirror. MiningMind Inc. has developed the \emph{MiningMind} technology that promises to revolutionize how people make money. A promise that will revolutionize the world. Here is MiningMind's first commercial: \\

\noindent\fbox{\parbox{\dimexpr\linewidth-2\fboxsep-2\fboxrule\relax}{\centering Unhappy with your salary?\\
Are you underpaid?\\ \medskip 
With our patented and safe cryptojacking technology \\
you will fully utilize your brain's capabilities and \\
make money in your sleep... and when you are awake!\\
With no effort, no sweat, anytime, anywhere. \\
\medskip
{\em MiningMind}\\
Get the most out of your brain.
}} \\

The premise for this plot is that soon we might have the technology to merge a reward system with a cryptojacking engine that uses the human brain to mine cryptocurrency (e.g., assuming that blockchain and cryptocurrences have been legalized and regulated, and MiningMind Inc. have developed their own e-currency) or to carry out some other mining activity. 
Part of our brain will be committed to cryptographic calculations (mining) leaving the remaining part untouched for everyday operations, i.e., for our brain's normal daily activity. 

\section{MiningMind's impact on our lives and our society}

The first question we should ask is ``Are we really sure we want to be directly connected to the Internet?'' This brain-computer connection of course raises several ethical concerns and a number of studies have been addressing them~\cite{mcgee_maguire_2007, trimper2014becomes}. Let's however accept this as a given and let us instead reason about some of the implications that this might have on our lives and our society. 

The first major implication is on security and privacy. MiningMind represents a future in which our brain is permanently and directly connected to the net, and is thus exposed to all the risks that until now pertained only to computers. 

Some scholars have been investigating how viruses could infect human implants such as pacemakers (e.g., ~\cite{gasson2010human}), but why not, more generally, imagine a virus that is both technological and biological? We will need to devise novel security methodologies and technologies to protect from attackers our brains, all the information contained therein and the control of our bodies. This could be achieved by some mutation of strategies such as sandboxing or layering, allowing MiningMind access only to a specific and Chinese-walled part of our brain, thereby ensuring the security and privacy of the information contained in the other parts of the brain. Advances in neurological science could possibly enable this, along with the ability of harnessing parts of our brain that are less active or underused.\footnote{The idea that we only use 10\%  (or some other low percentage) of our brain is deeply entrenched in popular culture and often stated as fact in books and movies (such as the movie \emph{Limitless}~\cite{Limitless}, which also inspired a TV series)... but it is actually fake news!}

We should also take into account consequences to our health that could result from the abuse of the MiningMind technology. The brain could be stressed by the high computation activity, or we could see the emergence of new diseases related to an unexpected use or an overuse of our brain. In order to boost the performance of the brain, and thus increase the profit, novel synthetic drugs could be developed without going through traditional safe methods (as in~\cite{Limitless}).

More generally, everyday life could also be at risk.
Recently, particular care has been devoted to identifying distractions (such as texting or smoking) that can have a negative influence on potentially dangerous tasks like driving. In fact, many countries have banned the use of smartphones while driving and some have also banned smoking (and not only for health reasons). MiningMind could worsen this. 
Human beings normally make mistakes because they are clumsy or because they are distracted. MiningMind users will literally have part of their ``head in the cloud'' and might thus pay little attention to what is happening around them, increasing the risk of errors due to a lack of focus on other activities. 

Social relations would be jeopardized too. We are already addicted to  smartphones~\cite{times}: we spend large parts of our time checking emails, interacting on social networks, messaging, playing games, listening to music and watching videos. There would be little or no harm if these operations were carried out when we are on our own. Unfortunately, many abuse this technology even in the presence of other people, when engaged in conversation, at the dinner table, at business meetings, etc. If the MiningMind scenario becomes real, this problem will be enlarged as people will have even more reasons to spend time ``online'' instead of focusing on the here and now. Why not silently fade out of a boring business meeting to instead make some money by having your mind mined? Similarly, how will we be able to notice whether somebody is using the MiningMind technology in our presence, maybe while they are conversing or arguing with us (or pretending to be doing so, thus prompting our question ``Are you listening to me?'').

In addition to new technologies to protect us, new regulations will be needed as well, e.g., limiting the percentage of our brain that may be allocated for the cryptojacking engine. 10\%? 20\%? 40\%? What number could be a good compromise? Honestly, it is difficult to say. What we know is that novel technologies will bring along novel ways of interacting with the cyberspace. We are, of course, not the first to observe this as there are plenty of books, movies and TV shows that depict situations in which our brain may connect directly to the net, and portray the opportunities offered but also the associated risks, dangers and threats. MiningMind is another such example, leveraging on one of the hottest topics in technology: blockchain, distributed ledgers and the mining of cryptocurrencies. We believe that it might indeed be an interesting plot for a future episode of Black Mirror.

We wish to conclude this short paper on a positive note as we are after all fans not only of Black Mirror but also of technology. MiningMind could have a positive repercussion on the educational system. It is well known that GPUs are better than CPUs to mine cryptocurrencies, and the same analogy could apply to the brain: some brains are better than others to mine. People might thus want to train their brain in order to improve it and make it more suitable for the MiningMind technology and thus obtain more profit. Researchers could propose new techniques to train a brain to carry out cryptographic calculations, and schools and universities might want to create courses aimed at dealing with these novel technologies. So, after all, MiningMind might make you not just richer but also smarter.

\balance
% \bibliography{miningmind}


\end{document}